\documentclass[10pt]{article}
\usepackage{amssymb,amsmath,cite,geometry,graphicx,url}
\long\def\@makefntext#1{\parindent 0cm\noindent
\hbox to 1em{\hss$^{\@thefnmark}$}#1}
\begin{document}

 \begin{center}
{\Large\bf
Reply to Wang and Unruh}\\  
\vspace{.2in}
{S.~C{\sc arlip}\\
       {\small\it Department of Physics}\\
       {\small\it University of California}\\
       {\small\it Davis, CA 95616, USA}\\
       {\small\it email: carlip@physics.ucdavis.edu}}
\end{center}
 \vspace*{1ex}
In \cite{Carlip1}, I suggested that a large cosmological constant migt be ``hidden''
in Planck-scale fluctuations of geometry and topology.  The thrust of the  paper was 
to show the existence of a large class of  initial data  for which the  expansion and 
shear, averaged over small regions, vanished even in the presence of an arbitrary 
cosmological constant.   Whether this behavior was preserved by evolution
was, I argued, fundamentally a question for quantum gravity.  As a first small step,
though, I showed that for short times, a classical time-slicing could be chosen for
which the average expansion $\langle K\rangle$ remained zero.  (Angle brackets 
$\langle\,\cdot\,\rangle$  will denote spatial averages over regions larger than Planck 
size but small compared to observed sizes.)   The  ``short times'' condition was 
necessary---the geometries considered in \cite{Carlip1}  contain high densities of 
marginally outer trapped surfaces, and the classical evolution quickly 
becomes singular \cite{Burkhartb}.
 
 The present Comment \cite{Wang} focuses on the question of classical
 evolution, but now concentrating on long-time behavior.  This is an interesting
 problem, of course, particularly if one has a way to resolve the singularities.
 For $\Lambda<0$, Wang and Unruh make a proposal for the treatment of certain 
``cosmological'' singularities in \cite{Wangb}, based on a construction rather different 
from  \cite{Carlip1}.
  
 Wang and Unruh first observe, correctly, that in the slicings of \cite{Carlip1} for which 
 $\langle K\rangle$ vanishes, the volume $V_{\mathcal{U}}$ of a fixed region remains
 time-dependent.  It's not obvious which criterion is more relevant: red shift observations
 measure something closer to expansion rate than volume, and it's unclear how one 
 would demarcate a ``fixed region  of space'' to compare volumes at different times.  
 But it is straightforward to repeat the construction of \cite{Carlip1}  to find a slicing 
 in which $V_{\mathcal{U}}$ is constant.  Indeed, starting with eqns.\ (1)  and (2) of 
 \cite{Wang} and taking further derivatives, one finds that
\begin{equation}
\frac{d^{n+1}V_{\mathcal{U}}}{dt^{n+1}} 
   = \int_{\mathcal{U}}\left(K\frac{d^{n}N}{dt^{n}} + \dots\right)\sqrt{g}\,d^3x \, .
 \label{a1}
\end{equation}
Since the derivatives of the lapse $N$ can be chosen independently on the initial 
hypersurface, it is easy to make the right-hand side of (\ref{a1}) vanish.  Of course,
as in \cite{Carlip1}, this is only a short-time condition, since singularities will
inevitably develop.
 
 Wang and Unruh next argue that the time-slicing of \cite{Carlip1} ``slow[s] down the
 progress of time'' in some areas and speeds it up in others, where by ``time'' they
 mean the proper time of a congruence of geodesic observers who start at rest on
 the initial hypersurface.\footnote{It's not obvious why these observers are important, 
 since the choice of initial hypersurface in \cite{Carlip1} is arbitrary.}  But this is 
 true of almost every  slicing used in general relativity: constant mean curvature slicing, 
 for instance, or maximal slicing, or the static slicing for any 
 inhomogeneous static  spacetime.  The authors argue for a special role for a slicing 
 corresponding Gaussian normal coordinates (lapse $N=1$, shift $N^i=0$).  But while 
 this choice can be useful,  it can also be misleading.  For the exterior Schwarzschild 
 solution, for instance,  such ``Novikov coordinates'' \cite{MTW} lead to time-dependent
 metric components, disguising the staticity.  This behavior  occurs more generally for 
 static inhomogeneous spacetimes, where it reflects the fact  that proper time depends 
 on the gravitational potential.  Even in cosmology, where a Gaussian normal slicing
 is common, it is typically only used for average properties, in an approximation
 in which local variations of the gravitational field are unimportant. 
  
 In the final section, Wang and Unruh show that under certain special
 circumstances, volumes in a spacetime with positive $\Lambda$   ``disastrously expand.''  
 Their conclusion is again correct, but the proof requires two conditions on the time-slicing: 
 that it is Gaussian normal, and that the average spatial scalar curvature $\langle{}^{(3)\!}R\rangle$ 
 vanishes.  These conditions greatly overdetermine the  slicing;  no such choice exists 
 for many spacetimes, and there is no  reason to expect it to exist for the 
 geometries discussed in \cite{Carlip1}.  For those geometries, it may  be that there is 
 \emph{some} slicing in which $\langle{}^{(3)}\!R\rangle=0$---this was left as an open 
 question---but there is no reason to think such a slicing  should be Gaussian normal.
 
The asymptotic behavior of vacuum spacetimes with a cosmological  constant is an important 
open question, but it is far from being settled.  There  are a few rigorous partial results 
in the literature \cite{Kleban,Mirbabayi,Moncrief,Creminelli}, but  these all require a Cauchy 
surface that either has constant mean curvature or is at least ``everywhere expanding.''   
The spacetimes of \cite{Carlip1} do not satisfy the first condition  \cite{Chruscielb}, and it 
seems quite unlikely  that they satisfy the second.  There is, I believe,
good reason to expect the answer to be complicated.  We know that a lattice of black 
holes in de Sitter space acts on a large scale very much like homogeneous
pressureless matter \cite{Yoo}, and that a stochastic background of gravitational radiation 
acts very much like a homogeneous radiation fluid \cite{Isaacson}.  But a universe
with a cosmological constant, matter, and radiation can exhibit many different behaviors:
it can expand, collapse, ``loiter,'' or, for finely tuned initial data, remain static.  When one
adds the complication of quantum fluctuations, which may (or may not) reproduce the
rich structure of the initial data of \cite{Carlip1}, there is no reason to expect a simple result.
 
\begin{flushleft}
\large\bf Acknowledgments
\end{flushleft}

This work was supported in part by Department of Energy grant DE-FG02-91ER40674.

\end{document}